\documentclass[twocolumn,twoside,slac_two]{revtex4}
\usepackage{graphicx}
\usepackage{fancyhdr}
\begin{document}

\title{CP Violation in the B$^0_s$ system}


\author{S. Donati}
\affiliation{University and INFN of Pisa, Largo Pontecorvo 2, 56127 Pisa, Italy}

\begin{abstract}
In this paper the most recent Tevatron results concerning 
CP violation in the $B_s$ system are reviewed. These are
the measurement of the direct CP asymmetry in the
$B^0_s\rightarrow K^-\pi^+$ decay performed by CDF
and the measurement of $\Delta\Gamma_s$ and $\phi_s$
performed by D0 in the $B^0_s\rightarrow J/\psi\phi$ decay.
\end{abstract}

\maketitle

\thispagestyle{fancy}


\section{Introduction}
One of the challenges for elementary particle physics is to
trace all possible sources of the violation of CP symmetry.
In the standard model, CP symmetry is violated through the
CKM mechanism. Although the standard model picture of CP
violation has so far been confirmed by all laboratory
measurements, the level of CP violation in the standard
model is too small to produce the observed baryon number
density in the universe. One source of CP violation arises
in the mixing of doublets of neutral mesons, as the $K^0$,
composed of a down quark and a strange quark, and the $B^0_s$
mesons, with the down quark replaced by a bottom quark.
This makes searching for CP violation in the $B^0_s$ decays
very interesting.
B hadrons are abundantly produced at the Tevatron Collider, 
the measured B$^+$ cross section is 2.78$\pm$0.24~$\mu$b
in the region of transverse momentum $p_T(B^+) >$~6.0~GeV/c 
and rapidity $\mid \eta(B^+)\mid <$~1~\cite{Bcrosssection}. 
This cross section is three orders of magnitude larger than 
at $e^+ e^-$ machines
running at the $\Upsilon (4S)$ and the available energy allows 
the production of the heavier $B^0_s$ (and $B_c$ and $\Lambda_b$) 
hadrons, although reduced by a factor $f_s/f_d(f_u)$ with respect to
$B^0(B^+)$. The challenge is extracting the interesting $B^0_s$ 
signals from a level of background which is three orders of
magnitude higher at production. This is achieved at CDF~II and
D0 with dedicated detectors, triggers and sophisticated analyses.

\section{The Tevatron Collider and the CDF~II and D0 Detectors}
The Tevatron Collider collides 36 $p\overline{p}$ bunches
at $\sqrt{s}$ = 1.96 TeV.  The design
instantaneous luminosity was 10$^{32}$ cm$^{-2}$s$^{-1}$ but 
the Tevatron now exceeds it and set the peak luminosity record 
at 2.8$\times 10^{32}$ cm$^{-2}$s$^{-1}$. With an already
integrated luminosity of 2.5~fb$^{-1}$, the
expectation is to have integrated $\sim$8~fb$^{-1}$ 
by the year 2009.

\subsection{The CDF~II Detector and Trigger}
\begin{figure}[ht]
\centering
\includegraphics[width=80mm]{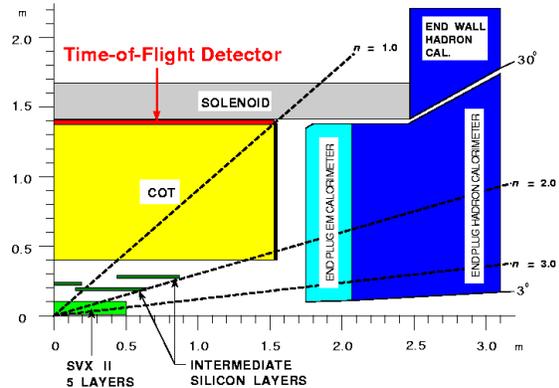}
\caption{Quarter view of the CDF~II tracker. Starting from the
beamline are visible the silicon detectors (L00, SVXII and ISL),
the drift chamber (COT), the Time of Flight detector and the solenoid.
Also reported are the plug calorimeters.
The central calorimeter and the muon chambers which surround the
solenoid are not reported in this drawing.} 
\label{fig:cdf_tracker}
\end{figure}

The elements of the CDF~II detector (Fig. \ref{fig:cdf_tracker}) 
most relevant for B physics analyses are the tracker, the

\noindent
particle identification detector and the muon system.
The CDF~II tracker is located within a 14.1 kG solenoidal magnetic
field and it is composed of a drift chamber and of silicon detectors.
There are three independent silicon detectors, SVXII, ISL and
L00, for a total of eight silicon layers, 704 ladders and 722,432 
channels~\cite{svxii}, posed between the radii of 1.5 cm and of 
28 cm from the beamline.
The Central Outer Chamber (COT, \cite{cot}) is located outside 
the silicon detectors and inside the time-of-flight detector
scintillators. The active volume of the COT spans 310~cm
in the beam direction, 43.4~cm and 132.3~cm in radius,
and the entire azimuth. The COT contains 30,240 sense wires
that run the length of the chamber between two end plates.
Approximately half of the wires are axial (run along the
$z$ direction) and half are small angle ($\pm 2^{\circ}$) stereo.
The $r-\phi$ view provides information for the $p_T$ measurement,
the $r-z$ view for the $\eta$ measurement.
The achieved performance of the integrated CDF~II tracker
is a transverse momentum 
resolution $\sigma (p_T)/p_T^2$ = 0.15~\% (GeV/c)$^{-1}$
and an impact parameter resolution
$\sigma (d)$ = 35 $\mu$m @2~GeV/c. This performance is
crucial for the B physics analyses.

CDF~II uses two complementary techniques
for particle identification, one is the dE/dx measurement 
in the COT, the other one is the time-of-flight measurement
in a dedicated detector.
The COT readout electronics allows to measure 
the pulse width, which is related to the amount of charge 
collected by the wire. The truncated mean (80~\%) computed on 
the hits
associated to a track provides a measurement of the specific
ionisation (dE/dx) in the chamber. A detailed calibration of
the dE/dx measurement has been performed using 
samples of kaons and pions 
from $D^{*+}\rightarrow D^0 \pi^+\rightarrow [K^-\pi^+]\pi^+$,
protons from $\Lambda^0\rightarrow p \pi^-$, and muons
and electrons from $J/\psi\rightarrow\mu^+\mu^-$ and
$J/\psi\rightarrow e^+ e^-$. The achieved $K/\pi$ separation
for $p_T>$~2~GeV/c is 1.4$\sigma$.
The Time-of-Flight detector (TOF, \cite{tof}) is composed of
216 scintillator bars istalled between the drift chamber and 
the solenoid magnet at a radius of roughly 138~cm. 
The time resolution on the single hit is 110~ps and the $K/\pi$
separation is better than 2$\sigma$ for $p_T <$~1.5~GeV/c.
By combining the dE/dx and the time-of-flight measurements,
the achieved $K/\pi$ separation is better tha 1.4~$\sigma$ 
in the entire momentum range.

The CDF~II central muon detector \cite{cmu} is located around 
the outside of the central calorimeter, which is 5.5 interaction
lengths thick, at a radius of 347~cm from the beam axis. 
The pseudorapidity coverage of the muon detector is $\mid\eta\mid <$~1.

CDF~II uses a three-level system to reduce the 1.7~MHz bunch 
crossing rate to 100~Hz written on tape. The Level 1 is a 
deadtimeless 7.6 MHz synchronous pipeline with 42 cells, 
which allows 5.5~$\mu$s to form a trigger decision. The
maximum sustainable Level 1 output rate is ~30~kHz.
The Level 2 is an asynchronous pipeline with an average
latency of 20~$\mu$s. While the events accepted by Level~1
are being processed by Level 2 processors, they are also 
stored in one of the four Level 2 buffers, waiting for 
Level~2 trigger decision. Each buffer is emptied 
when the Level 2 decision for the corresponding event 
has been asserted: if the event has been accepted, the
buffer is read out, else it is simply cleared. 
If the Level~2 trigger decision takes 
too much time and the four buffers are all filled, 
the Level~1 accept is inhibited. This is a source 
of deadtime for the CDF~II trigger. The maximum 
Level 2 output rate is 300~Hz. The Level 3 trigger
is made of a CPU farm and has a maximum output rate
of 100~Hz.

The heart of the Level 1 trigger is the eXtremely
Fast Tracker (XFT, \cite{XFT}), the trigger track
processor that identifies high transverse momentum
($p_T> 1.5$~GeV/c) charged tracks in the transverse
plane of the COT. 
The XFT tracks are also extrapolated to the calorimeter
and to the muon chambers to generate electron and muon
trigger candidates. 

The Online Silicon Vertex Tracker (SVT, \cite{SVT}) is part
of the Level 2 trigger. It receives the list of XFT tracks
and the digitised pulse heights on the axial layers of the
silicon vertex detector. The SVT links the XFT tracks to the 
silicon hits and reconstructs tracks with offline-like quality.
In particular the resolution on the impact parameter, which is
a crucial parameter to select B events since they tipically 
show secondary vertices, is 35~$\mu$m for 2~GeV/c tracks. 
The SVT efficiency is ~85~\% per track.

The Level 3 trigger is implemented on a CPU farm which allows 
to perform an almost offline-quality event reconstruction.

CDF~II has basically three families of triggers for B physics:
the dimuon trigger, the semileptonic trigger and the hadronic 
trigger. 
The dimuon trigger selects muon pairs with
transverse momenum as low as 1.5~GeV/c. It is mostly used to
select $J/\psi$s and $\psi(2S)$, to reconstruct the many decay
modes of the B hadrons ($B^0$, $B^+$, $B^0_s$, $B_c$, and $\Lambda_b$)
containing a $J/\psi$ decaying to muon pairs, and to select 
$\Upsilon\rightarrow \mu^+\mu^-$ decays, or muon 
pairs for the search of the rare $B\rightarrow \mu^+\mu^- X$ decays,
or for $b\overline{b}$ correlation studies.
The semileptonic trigger selects events with a lepton ($\mu$ or $e$)
with $p_T~>$~4 GeV/c and an SVT track with $p_T >$~2~GeV/c and impact
parameter above 120~$\mu$m. 
The hadronic trigger selects track pairs with $p_T > 2$~GeV/c
and $p_{T1}+p_{T2} >$ 5.5~GeV/c, with an opening angle in the
transverse plane below 135$^{\circ}$,
impact parameter above 100~$\mu$m,
a decay length above 200~$\mu$m. For the two-body decay
trigger path, optimised to collect $B\rightarrow h^+h^{'-}$ decays, 
the track pair is requested to point back to the
primary vertex, by requiring that the impact parameter of the 
reconstructed B is below 140~$\mu$m. To select hadronic multibody
decays, like $B^0_s\rightarrow D^-_s\pi^+$, the request on the
pointing back to the primary vertex has low efficiency, since the
track pair provides only a partial reconstruction of the multibody
decay, and it is not applied.

\subsection{The D0 Detector and Trigger}
\begin{figure}[ht]
\centering
\includegraphics[width=80mm]{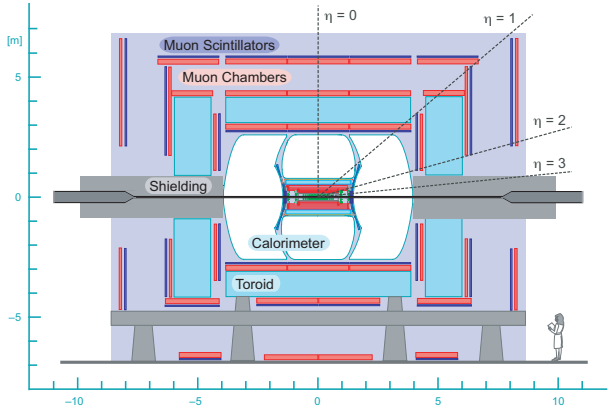}
\caption{View of the D0 detector.} 
\label{fig:d0}
\end{figure}

The D0 detector (Fig.~\ref{fig:d0}) consists of a magnetic 
central-tracking system, comprised of a silicon
microstrip tracker and a central fiber
tracker, both located within a 2~T superconducting magnet.
The fiber tracker has eight thin coaxial barrels, each supporting
two doublets of overlapping scintillating fibers, one doublet being 
parallel to the collision axis, and the other alternating by
$\pm 3^{\circ}$ relative to the axis.
Central and forward preshower detectors located just outside of 
the superconducting coil in front of the calorimetry are constructed
of several layers of extruded triangolar scintillator strips.
The nextr layer of detection involves three liquid-argon/uranium
calorimeters: a central section covering $\mid\eta\mid < 1.1$ and
two endcap calorimeters that extend coverage to 
$\mid\eta\mid< 4.2$ all housed in separate cryostats.
The muon system is located beyond the calorimetry, and consists
of a layer of tracking detectors and scintillator trigger 
counters before 1.8~T iron toroids, followed by two similar

\noindent
layers after the toroids.
A muon originating in a $p\overline{p}$ collision traverses
the silicon detector and the scintillating fiber tracker in 
the 2~T solenoidal magnetic field, the calorimeter, and the
muon spectrometer and 1.8~T magnetised iron toroids.
The momentum of the muon is measured twice: once by the
local muon system and once by the central-tracking system.
The polarities of the toroid and the solenoid magnetic 
fields are reversed roughly every two weeks so that the four
solenoid-toroid polarity combinations are exposed to approximately
the same integrated luminosity. This allows cancellation of first-order
effects of the detector geometry. 

The trigger and data acquisition systems are designed to accomodate
the high luminosities of run~II. Based on preliminary information
from tacking, calorimetry, and muon systems, the output of the first 
level of the trigger is used to limit the rate for accepted events
to 2~kHz. At the next trigger stage, with more refined information,
the rate is reduced further to 1~kHz. The third and final level of
the trigger, with access to all the event information, uses software
algorithms and a computing farm, and reduces the output rate to 
50~Hz, which is written tape. B triggers are based on single letpons
($p_T > 3.0$~GeV/c) and on muon pairs ($p_t > 2.0$~GeV/c).

\begin{figure}[ht]
\centering
\includegraphics[width=80mm]{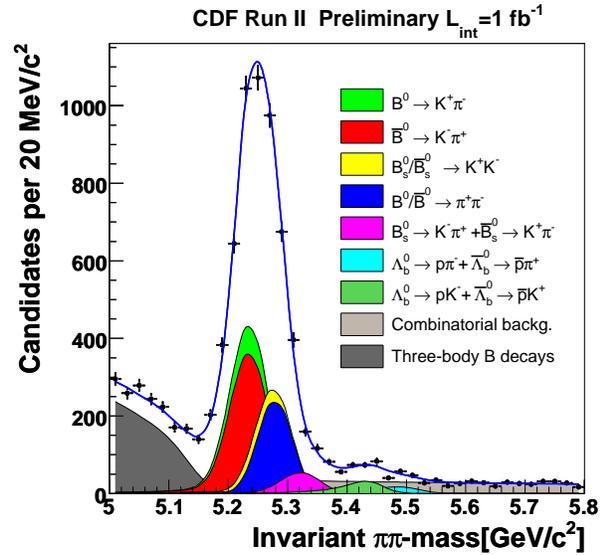}
\caption{Invariant mass distribution of
$B^0_{(s)}\rightarrow h^+ h^{'-}$ candidates 
passing all selection requirements. The invariant
mass is computed by assigning the pion mass to both
tracks in the decay.} 
\label{fig:b0hh}
\end{figure}

\section{CDF Measurement of Direct CP Asimmetry in the
$B^0_s\rightarrow K^-\pi^+$ Decay}

CDF~II reported the first evidence of the decay and measurement of the direct
CP asimmetry in the $B^0_{(s)}\rightarrow K^-\pi^+$ decay reconstructed
in the data taken by the hadronic trigger (~1~fb$^{-1}$).
In the offline analysis an unbiased optimisation procedure determined 
a tightened selection on track-pairs fit to a common decay vertex,
optimising the sensitivity for discovery and limit setting \cite{punzi}.
In addition to tightening the trigger cuts, the discriminating power
of the $B^0_s$ meson isolation and of the information provided by the
3D reconstruction capability of the CDF tracking were used, allowing 
a great improvement in the signal purity. The resulting $\pi\pi$-mass
distribution (Fig.~\ref{fig:b0hh}) shows a clean signal of 
$B^0_{(s)}\rightarrow h^+ h^{'-}$ decays. In spite of a good mass
resolution ($\approx 22$~MeV/c$^2$), the various 
$B^0_{(s)}\rightarrow h^+ h^{'-}$ mdoes overlap into an unresolved
mass peak.

\begin{figure}[ht]
\centering
\includegraphics[width=80mm]{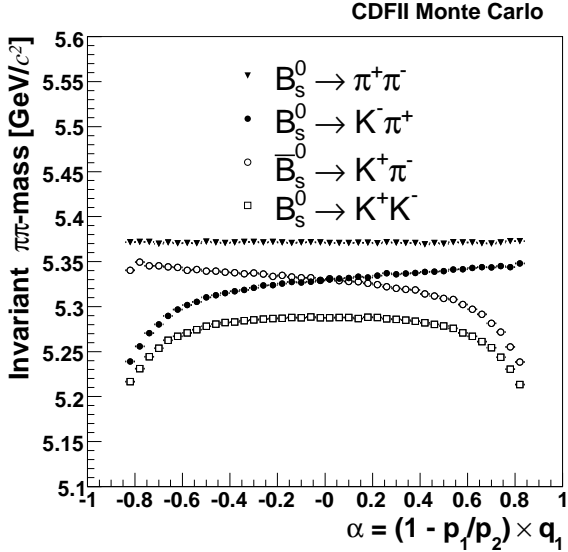}
\caption{$M_{\pi\pi}$ versus signed momentum imbalance
$\alpha = (1-p_1/p_2)q_1$ for the $B^0_s\rightarrow h^+ h^{'-}$
decays. Similar curves are for the 
$B^0_d\rightarrow h^+ h^{'-}$
and $\Lambda^0_b\rightarrow ph$ decays} 
\label{fig:b0s}
\end{figure}

The resolution in invariant mass and in particle identification is
not sufficient for separating the individual decay modes on an event
by event basis, therefore an unbinned maximum likelihood fit was
performed, combining kinematic and particle identification information,
to statistically determine both the contribution of each mode, and
the relative contributions to the CP asimmetries.
For the kinematic portion, three loosely correlated variables were used
to summarise the information carried by all possible values of invariant 
mass of the B candidate, resulting from different mass assignments to the 
two outgoing particles. They are: 
(a) the mass $M_{\pi\pi}$ calculated
with the charged pion mass assignment to both particles;
(b) the signed momentum imbalance $\alpha = (1-p_1/p_2)q_1$,
where $p_1 (p_2)$ is the lower (higher) of the particle
momenta, and $q_1$ is the sign of the charge of the particle
momentum $p_1$ (Fig. \ref{fig:b0s}); 
(c) the scalar sum of the particle momenta $p_{tot} = p_1 + p_2$.
Particle identification information derives mostly from the
dE/dx measurement in the drift chamber which provides a 1.4$\sigma$
separation between pions and kaons in the momentum range of interest.
A sample of 1.5M $D^{*+}\rightarrow D^0\pi^+\rightarrow [K^-\pi^+]\pi^+$
decays, where the $D^0$

\begin{figure}[ht]
\centering
\includegraphics[width=80mm]{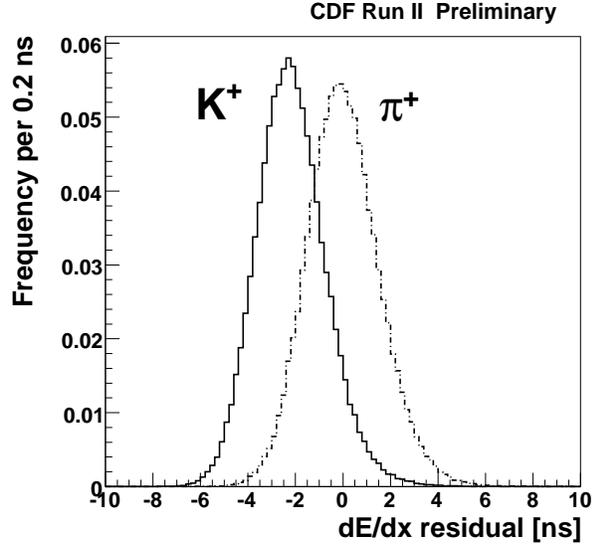}
\caption{Tagged $D^0\rightarrow K^-\pi^+$ decays from
 $D^{*+}\rightarrow D^0\pi^+\rightarrow [K^-\pi^+]\pi^+$:
distribution of dE/dx around the average pion response,
for calibration samples of kaons and pions.} 
\label{fig:residuals}
\end{figure}

$D^{*+}\rightarrow D^0\pi^+\rightarrow [K^-\pi^+]\pi^+$
decays, where the $D^0$ 
de\-cay products are identified by the charge 
of the $D^{*+}$ pion, was used to calibrate the dE/dx response over
the tracking volume and over time (Fig. \ref{fig:residuals}).
The mass resolution function was parameterised using the detailed
detector simulation. To take into account non-Gaussian tails due 
to the emission of photons in the final state, soft photon emission
was included in the simulation using recent QED calculations
\cite{baracchini}. 
The quality of the mass resolution was verified
using about 500k $D^0\rightarrow K^- \pi^+$ decays and improved 
agreement was found between data and the simulation including the
final state radiation.

\begin{figure}[ht]
\centering
\includegraphics[width=80mm]{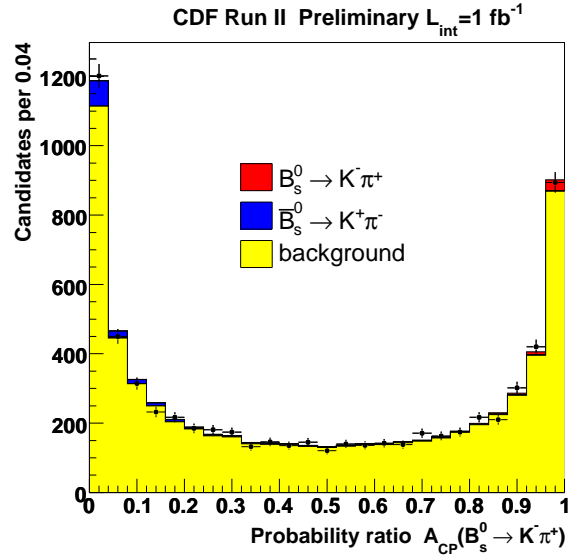}
\caption{Likelihood ratio for the measurement of $A_{CP}$
in the $B^0_s\rightarrow K^-\pi^+$ decay.} 
\label{fig:lkhdratio}
\end{figure}

In addition to the improved measurement of branching fractions 
of the already known modes ($B^0\rightarrow\pi^+\pi^-$, 
$B^0\rightarrow K^+\pi^-$ and $B^0_s\rightarrow K^+ K^-$)
and of the direct CP asymmetry $A_{CP}(B^0\rightarrow K^+\pi^-)$,
the analysis observes three new rare modes for the first time
($B^0_s\rightarrow K^-\pi^+$, $\Lambda^0_b\rightarrow p\pi$
and $\Lambda^0_b\rightarrow pK$), with a significance 
respectively of 8.2$\sigma$, 6.0$\sigma$ and 11.5$\sigma$,
which includes both statistical and systematic undertainty.
To convert the yields returned from the fit into relative
branching fractions, we applied corrections for efficiencies
of trigger and offline selection requirements for different 
decay modes.

The relative efficiency corrections between modes
do not exceed 20~\%.
Most corrections were determined from the
detailed detector simulation, with some exceptions which were
measured using the data. The only correction needed for the 
direct CP asymmetries $A_{CP}(B^0\rightarrow K^+\pi^-)$ and
$A_{CP}(B^0_s\rightarrow K^-\pi^+)$ was a $<0.6$~\% shift due
to the different probability for a $K^+$ and $K^-$ to interact
with the tracker material. The measurement of this correction
has been performed using a sample of 1M of prompt 
$D^0\rightarrow K^-\pi^+$ decays reconstructed and selected
using the same criteria as 
$B^0_{(s)}\rightarrow h^+ h^{'-}$ decays.
Assuming the standard model expectation of
$A_{CP}(D^0\rightarrow K^-\pi^+)=0$, the difference between
the number of reconstructed $D^0\rightarrow K^-\pi^+$ and
$\overline{D}^0\rightarrow K^+\pi^-$ provides a measurement
of the detector induced asymmetry between $K^+\pi^-$ and
$K^-\pi^+$.
The measured branching fraction of the newly observed mode
$BR(B^0_s\rightarrow K^-\pi^+) = (5.0 \pm 0.75 \pm 1.0)\times 10^{-6}$
is in agreement with the latest theoretical predictions
\cite{b0skpi1}, which are lower than previous predictions
\cite{b0skpi2}.

The analysis measures for the first time the direct CP asymmetry 
$A_{CP}(B^0_s\rightarrow K^-\pi^+) = 0.39 \pm 0.15 \pm 0.08$, which
favors a large CP violation in the $B^0_s$ decays. A robust test
of the standard model can be performed by comparing the measurements
of the direct CP asymmetry in the $B^0_s\rightarrow K^-\pi^+$ and
$B^0\rightarrow K^+\pi^-$ decays \cite{smtest}. The estimated 
expected value for $A_{CP}(B^0_s\rightarrow K^-\pi^+) \approx 0.37$
is in agreement with the CDF~II measurement.
The branching fraction $BR(B^0_s\rightarrow K^+ K^-) = 
(24.4 \pm 1.4 \pm 4.6) \times 10^{-6}$ is in agreementt with
the latest theoretical expectations \cite{bskk}. With a large 
$B^0_s\rightarrow K^+ K^-$ sample ($\approx$ 1300 ev/fb$^{-1}$)
and the flavor taggers optimised and $x_s$ measured, CDF has all
the ingredients for a time dependent $A_{CP}$ measurement in
this mode \cite{fleischer}, which is particularly interesting 
with the full statistics (8 fb$^{-1}$) expected for run~II.

\section{Measurements of $\Delta\Gamma_s$ and $\phi_s$}
The standard model predicts sizeable mass and decay width 
differences between the light and heavy eigenstates of the
mixed $B^0_s$ system. The CP violating phase is expected
to be small, thus the two mass eigenstates are expected to
be CP eigenstates. Phenomena beyond the standard model may
reduce the observed $\Delta\Gamma$ compared to the standard
model prediction 
$\Delta\Gamma\approx\Delta\Gamma_{SM}\times cos(\phi_s)$
\cite{phis}.
The decay $B^0_s\rightarrow J/\psi\phi$ gives rise to both 
CP-even and CP-odd final states, which can be separated 
through a study of the time-dependent angular distribution
of the decay products of the $J/\psi$ and $\phi$ mesons.
This allows to measure the lifetime difference between the
two states and provides sensitivity to the mixing phase
through the interference terms between the CP-even and
CP-odd waves \cite{fleischer1}. The angular analysis of the 
$B^0_s\rightarrow J/\psi\phi$ decays has been performed 
both by CDF and D0. Both experiments reconstruct the decay
modes in samples collected by the muon triggers, which 
require the presence of muon segments in the muon detectors
matched to central tracks.
In the offline analysis $J/\psi$ and $\phi$ candidates are
required to be consistent with coming from a common vertex
and to have an invariant mass compatible with the $B^0$ mass
The proper decay length $ct$ is defined by the relation
$ct = L^{B}_{xy}\cdot M_{B^0_s}/p_T$, where $M_{B^0_s}$ is 
the measured mass of the $B^0_s$ candidate.
A simultaneous unbinned maximum likelihood fit is performed
to the mass, proper decay length and three decay angles.
In the coordinate system of the $J/\psi$ rest frame, where
the $\phi$ meson moves in the $x$ direction, the $z$ axis
is perpendicular to the decay plane of $\phi\rightarrow K^+ K^-$
and $p_y(K^+)>0$, the transversity polar and azimuthal angles
($\theta$, $\phi$) describe the direction of

\begin{figure}[ht]
\centering
\includegraphics[width=80mm]{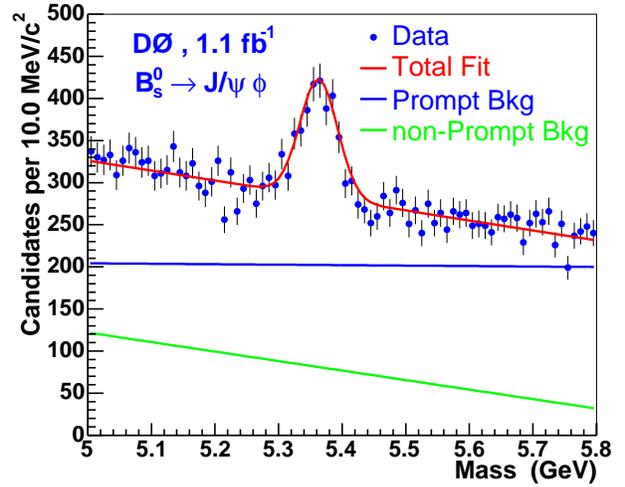}
\caption{Invariant mass distribution of the ($J/\psi$, $\phi$)
system for the $B^0_s$ candidates. The curves are projections
of the maximum likelihood fit.} 
\label{fig:d001}
\end{figure}

\noindent
the $\mu^+$, and $\psi$ between $\vec{p}(K^+)$ and $-\vec{p}(J/\psi)$
in the $\phi$ rest frame.
From the analysis of 1.1~fb$^{-1}$ of data D0 estimates
a yield of 1039$\pm$45 $B^0_s$ events, with an average
lifetime $\tau (B^0_s) = 1.52 \pm 0.08(stat)\pm 0.03(syst)$~ps
and a width difference between the two mass eigenstates
$\Delta\Gamma = 0.12^{+0.08}_{-0.10}(stat)\pm 0.02(syst)$~ps$^{-1}$.
Allowing for CP violation in the $B^0_s$ mixing the first
direct constraint on the CP violating phase
$\phi_s = -0.79 \pm 0.56(stat) \pm 0.14 (syst)$ 
is obtained \cite{prl98121801}.
Using a sample of 355~pb$^{-1}$ of data CDF reconstructed
203$\pm$15 $B^0_s$ decays and with an angular analysis measured 
$\tau(B^0_s) = 1.40^{+0.15}_{-0.13}(stat)\pm 0.02(syst)$~ps and
$\Delta\Gamma = 0.47^{+0.19}_{-0.24}(stat)\pm 0.01(syst)$~ps$^{-1}$
\cite{cdfbsjpsiphi}. CDF is currently updating the analysis to 
1~fb$^{-1}$ of data. At the moment of writing these proceedings 
the CDF results are not public yet.

\subsection{D0 charge asymmetry measurement}
D0 has measured the dimuon charge asymmetry
$A = (N^{++}-N^{--})/(N^{++}+N^{--})$, where $N^{++}$ ($N^{--}$)
is the number of events with two positive (negative) muon candidates
passing selection cuts. This inclusive, tagged measurement is
sensitive to CP violation in $B^0_s$ mixing. The dimuon charge
asymmetry $A$ has contributions from both $B^0$ and $B^0_s$,
therefore this measurement at the Tevatron collider is complementary
to similar measurements at B factories that are sensistive only to
$A_{B^0}$, not  $A_{B^0_s}$. The D0 measurement is 
$A = -0.0092 \pm 0.0044(stat) \pm 0.0032(syst)$
\cite{d0prd}. There are several sources of systematic 
errors contributing to this measurement, such as the detector
effects, which are significantly reduced by averaging over
the samples collected with the different 

\begin{figure}[ht]
\centering
\includegraphics[width=80mm]{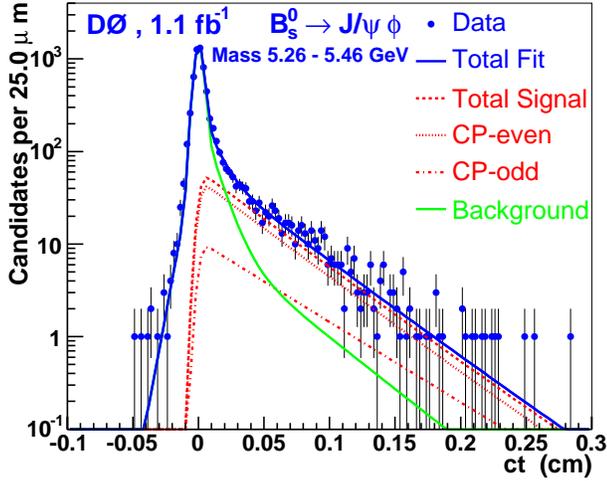}
\caption{Proper decay length $ct$ of the $B^0_s\rightarrow J/\psi\phi$ 
candidates in the signal mass region. The curves show the signal contribution,
the CP-even and CP-odd contributions of the signal, the background
and the total curve.} 
\label{fig:d002}
\end{figure}

\noindent
combinations of
magnetic field polarities, the different $K^{\pm}$ interaction
rates with the detector material which generates different
$K^{\pm}\rightarrow \mu^{\pm}$ rates, the presence of the
dimuon cosmic rays, and the hadronic punch-through.
Both $B^0_d$ and $B^0_s$ contribute to this quantity
($A = A_d + \alpha\cdot A_s$, where $A_d$ and $A_s$ are
respectively the charge asymmetries of the $B^0$ and 
$B^0_s$ semileptonic decays, and $\alpha$ is a coefficient
which depends on the $B^0$ and $B^0_s$ production rates
and mixing parameters, which can be computed using
the world average values 
$\alpha = 0.70 \pm 0.07(syst) \pm 0.10 (PDG)$).
The asymmetry $A_d$ has been measured at B factories
where only $B^0$ and $B^{\pm}$ are produced, and the
resulting average value is $A_d = -0.0047 \pm 0.0046$ \cite{hfag}.
This measurement can be used to extract $A_s$ from the measurement
of $A$ and the resulting value is $A_s = -0.0064 \pm 0.0101$, where
the statistical and systematic uncertainties are added in quadrature.

D0 has also measured a time integrated flavour untagged charge asymmetry
$A^{unt}_s$ in the semileptonic $B^0_s$ decays, defined as
$A^{unt}_s = [N(\mu^+ D^-_s)-N(\mu^- D^+_s)]/[N(\mu^+ D^-_s)+N(\mu^- D^+_s)]$
where the $\mu^{\pm}D^{\mp}_s$ pairs are produced in the semileptonic
$B^0_s$ decays. This asymmetry is called untagged since the initial
flavor of the $B^0_s$ meson is not determined. The measurement has
been performed in a dataset of about 
1.3~fb$^{-1}$ of integrated luminosity, using the decay
$B^0_s\rightarrow \mu D_s\nu X$, with $D_s\rightarrow \phi\pi$,
$\phi\rightarrow K^- K^+$.
$B^0_s$ candidates were reconstructed by clustering a muon
with $p_T>2.0$~GeV/c with three charged tracks, two of which 
assigned the kaon mass and required to have an invariant
mass (1.004 GeV/c$^2 < M(K^+ K^-) <$ 1.034 GeV/c$^2$) consistent
with that of a $\phi$ meson. The optimal selection on several
discriminating variables (the angle between the $D_s$ and 
the $K$ momenta in 

\begin{figure}[ht]
\centering
\includegraphics[width=85mm]{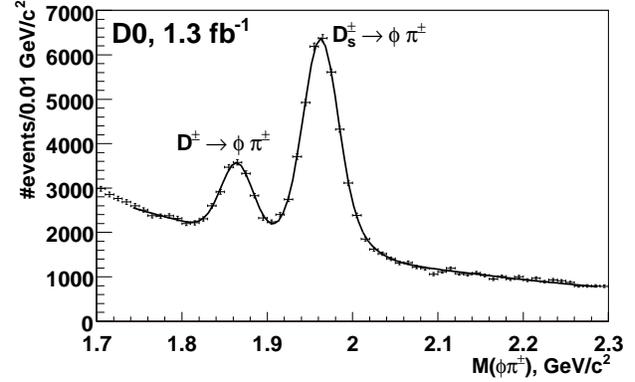}
\caption{Invariant mass distribution $M(\phi\pi)$ for the 
selected $B^0_s$ candidates. The curve shows the result
of a fit with a double gaussian function for the $D$ signals
and an exponential for the background.} 
\label{fig:d0semi}
\end{figure}

\noindent
the $K^+K^-$ center of mass frame, the
isolation of the ($\mu D_s$) system, the $\chi^2$ of the
$D_s$ vertex, the invariant masses $M(\mu D_s)$ and
$M(K^+ K^-)$ and $p_T(K^+ K^-)$) was determined by maximising
the predicted ratio $S/\sqrt{S+B}$. The resulting mass distribution
of the $D^{\pm}_s$ candidates is reported in Fig.~\ref{fig:d0semi}.
The measured value of the time integrated untagged charge 
asymmetry is $A^{unt}_s = [1.23 \pm 0.97(stat)\pm 0.17(syst)]\times 10^{-2}$.
The main systematics (added in quadrature in the final result) are
due to the different $\pi^{\pm}$ interaction with the material, the
estimate of the $B^0_s$ fraction in the $\mu D_s$ sample and the
fitting procedure (estimated by varying the masses and widths 
of the peaks and the slope of the background by 1$\sigma$).
Statistics dominates the error on this measurement and it
will be improved in the future with the increase of statistics
and addition of new decay modes \cite{d0semi}.

\begin{figure}[ht]
\centering
\includegraphics[width=80mm]{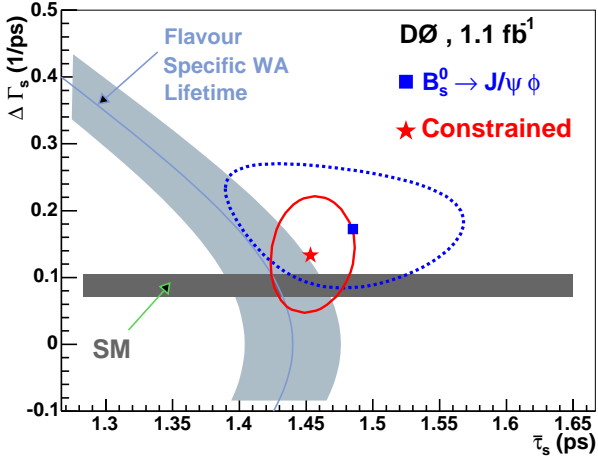}
\caption{Error ellipse in the $\Delta\Gamma_s$ versus $\tau_s$ 
plane for the fit to the $B^0_s\rightarrow J/\psi\phi$ 
data and for the fit with the constraint from the two D0 measurements
of the charge asymmetry in the semileptonic $B^0_s$ decay, and from 
the world average flavor specific lifetime. Also shown is the 1$\sigma$
band representing the world average result for $\tau_{fs}$ and the
1$\sigma$ band representing the theoretical prediction
$\Delta\Gamma_s = 0.088 \pm 0.017$~ps$^{-1}$ \cite{dgtheory}} 
\label{fig:d01}
\end{figure}

\begin{figure}[ht]
\centering
\includegraphics[width=90mm]{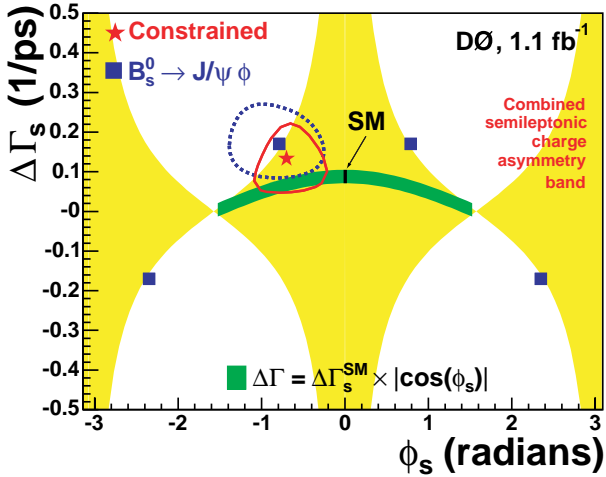}
\caption{Error ellipse in the $\Delta\Gamma_s$ versus $\phi_s$
plane for the fit to the $B^0_s\rightarrow J/\psi\phi$ data and 
for the fit with the constraint from the two D0 measurements of
the charge asymmetry in semileptonic $B^0_s$ decay and from 
teh world average flavor-specific lifetime. The central values
for all four solutions of the unconstrained fit are indicated
by blue squares. Also shown is the band representing the relation
$\Delta\Gamma_s = \Delta\Gamma^{SM}_s\times|cos(\phi_s)|$ with
$\Delta\Gamma^{SM}_{s} = 0.088\pm 0.017$~ps$^{-1}$.} 
\label{fig:d02}
\end{figure}

The two above D0 measurements are nearly independent as well
as the systematic uncertainties. Thus they can be combinated
to give the best estimate of the charge asymmetry in the
semileptonic $B^0_s$ decays $A_s = 0.0001 \pm 0.0090$.
$A_s$ can be related to the CP phase $\phi_s$ by the 
formula $A_s = (\Delta\Gamma_s/\Delta M_s)\cdot tan(\phi_s)$
\cite{beneke}. Using the CDF~II measurement 
$\Delta M_s = 17.8 \pm 0.1$~ps$^{-1}$ \cite{cdfxs} the following
constraint is obtained $\Delta\Gamma_s \cdot tan(\phi_s) = 
A_s\cdot\Delta M_s = 0.02 \pm 0.16$~ps$^{-1}$.
The fit to the $B^0_s\rightarrow J/\psi \phi$ data has been repeated
by D0 including the above constraint from the measuremnt of $A_s$
and the likelihood contours are presented in Fig.~\ref{fig:d01}
and Fig.~\ref{fig:d02},respectively in the $\Delta\Gamma_s$-$\tau_s$
plane and $\Delta\Gamma_s$-$\phi_s$ plane. There is an unresolved
4-fold ambiguity of the solutions. For the solution with $\phi_s<0$
and smaller absolute value,
the decay width and CP violating phase are measured to be
$\Delta\Gamma_s = 0.13\pm0.09$~ps$^{-1}$ and 
$\phi_s = -0.70^{+0.47}_{-0.39}$, consistent with the standard model
predictions. The measurement uncertainty is dominated by the limited 
statistics.

\section{Conclusions}
In this paper we have reviewed CDF and D0 results concerning
CP violation in the $B^0_s$ sector. Both collaborations have
been very active in this area. CDF has performed a measurement
of the direct CP asymmetry in the 
$B^0_s\rightarrow K^-\pi^+$ decay mode using the data collected
by the hadronic trigger. Both collaborations have performed 
angular analyses of the $B^0_s\rightarrow J/\psi\phi$ which
provide measurements of $\Delta\Gamma_s$ and $\phi_s$. D0
has performed charge asymmetry measurements in the single
muon and dimuon samples. With the 2.5~fb$^{-1}$ already on 
tape and the 8~fb$^{-1}$ expected by 2009 the CDF and D0
collaborations are confident to have many more and better
results soon. 

\bigskip % extra skip inserted

\end{document}